# Metasurface-Based Full-Parameter Optical Multiplexing

Rui Wei, Hongsheng Shi, Boyou Wang, Baojun Li* and Yanjun Bao*

Guangdong Provincial Key Laboratory of Nanophotonic Manipulation, Institute of Nanophotonics, College of Physics & Optoelectronic Engineering, Jinan University, Guangzhou 511443, China

*Corresponding Authors: Y. Bao (yanjunbao@jnu.edu.cn), B. Li (baojunli@jnu.edu.cn)

**Abstract:** Optical multiplexing is a key technique that enhances the capacity of optical systems by independently modulating various optical parameters to carry distinct information. Among these parameters, wavelength, polarization, and angle are the primary ones for multiplexing in plane waves with uniform cross-sectional distribution. While metasurfaces have recently emerged as a powerful platform for optical multiplexing, they are typically restricted to partial parameter multiplexing and exhibit a low number of multiplexing channels. In this work, we propose and experimentally demonstrate the full-parameter multiplexing of polarization, wavelength, and angle, achieving hundreds of distinct multiplexing channels—the largest reported to date. Our design utilizes a gradient-based optimization algorithm to enable high-efficiency performance and independent functionalities with minimal cross-talk among channels. This approach represents a significant advancement in metasurface design and optical multiplexing, with potential applications in complex and dynamic optical systems.

# Introduction

Light, the fundamental carrier of information in optical communications, possesses a myriad of parameters that define its characteristics, including amplitude, phase, polarization, wavelength, and angle of incidence/observation (or direction of wavevector). Optical multiplexing[1], a technique that leverages these diverse parameters of light, significantly increases the capacity and efficiency of optical communication systems, thereby accommodating the growing demand for higher data rates. Recently, metasurfaces, planar structures consisting of subwavelength-scale elements arranged in precise patterns, have emerged as a powerful platform offering unprecedented control over the properties of light[2-9]. These metasurfaces exhibit remarkable versatility in tailoring the various parameters of light, rendering them highly suitable for advanced optical multiplexing techniques[10-15].

For plane waves with uniform cross-sectional distribution, variations in initial amplitude and phase values do not introduce new information, making wavelength, polarization, and angle the three primary parameters for optical multiplexing. Numerous studies have demonstrated the potential of metasurfaces for various optical multiplexing applications, as illustrated in Fig. 1a. For angle multiplexing alone, the multiplexing number has increased from 2[16] to 4[17], 9[18], and eventually 25[19]. These techniques encode optical information into holograms or printed images based on incident or observed angles. For wavelength multiplexing, spatial multiplexing[20, 21] methods have been used to encode different optical holograms for different wavelengths. Polarization multiplexing typically utilizes its two orthogonal states, increasing the channel number to two[22-28]. The most common polarizations used are linear[22-24] and circular[25-28], where metasurface elements can act as linearly birefringent devices or induce an additional geometric phase to flexibly control the wavefronts. The examples above each involve only one multiplexing parameter. Multiplexing with multiple parameters becomes challenging due to the multi-objective nature of the problem. Recently, inverse design approaches employing algorithms like genetic algorithms[29], statistical machine learning[30], and adjoint methods[31] have been used to achieve two-orthogonal-polarization and multiple-wavelength multiplexing channels. However, these works typically exhibit a low number of multiplexing channels and have not yet realized multiplexing of all three parameters.

In this study, we propose and experimentally demonstrate the multiplexing of all optical parameters of plane waves: polarization, wavelength, and angle of observation, as illustrated by the red pentacle in Fig. 1a. This sets a new benchmark for the number of multiplexing channels, comprising two polarizations, three wavelengths, and over 25 observed angles, resulting in a total of 150 multiplexing channels—the largest reported to date, to our knowledge. To achieve such functionalities, we developed a gradient-based optimization algorithm to design the metasurface, enabling independent functionalities under different combinations of the three optical parameters with minimal cross-talk among channels. Our work represents a significant advancement in the field of metasurface design and optical multiplexing.

## Results and Discussions

Figure 1b presents a schematic of our designed metasurface, capable of displaying independent optical images under combinations of $N$ wavelengths ($\lambda_1, \lambda_2, ..., \lambda_N$), two polarizations [left-circular polarization (LCP) and right-circular polarization (RCP)], and $M \times M$ different observed angles ($\theta_{xi}, \theta_{yj}$, where i, j = 1,2,…$M$), resulting in a total of $2M^2N$ multiplexing channels. The metasurface design utilizes a nanoblock structure as the basic unit, characterized by three geometric parameters: length ($W_1$), width ($W_2$), and rotational angle ($\theta$). These elements are described by the Jones matrix of a conventional, linearly birefringent wave plate:

$$J(\lambda_n) = R(-\theta) \begin{bmatrix} A(\lambda_n) & 0 \\ 0 & B(\lambda_n) \end{bmatrix} R(\theta) \tag{1}$$

where $A$ and $B$ represent the complex-amplitude coefficients of light linearly polarized along the fast and slow axes, respectively, both of which are wavelength-dependent, providing the degree of freedom for wavelength multiplexing. When considering only circularly polarized incidence and cross-polarized transmission, the transmitted field coefficients are (see Supplementary Information):

$$C^k = \frac{A-B}{2} e^{i2\sigma_k \theta} \quad (k=1, 2) \tag{2}$$

where $\sigma_k = \pm 1$, with the sign determined by the incident polarization states.

Before implementing the full-parameter multiplexing, we first consider the single-wavelength case with only polarization and angle multiplexing. In this scenario, the coefficients $A$ and $B$ can be represented by pure phase modulation with $A = e^{i\varphi_x}$ and $B = e^{i\varphi_y}$. We then have $C^k = iDe^{i\alpha_k}$ with $D = \sin\left[(\varphi_x - \varphi_y)/2\right]$ and $\alpha_k = (\varphi_x + \varphi_y)/2 + 2\sigma_k \theta$ (see supplementary information), indicating that the amplitude is the same under both polarization channels (Fig. 2a). The two phases $\varphi_x$ and $\varphi_y$ are independent and can cover the full $2\pi$ range by sweeping the geometric width and length of the nanoblock. With the additional degree of freedom of $\theta$, the parameters $D$, $\alpha_1$, and $\alpha_2$ can take independent and arbitrary values.

With an arbitrary complex amplitude distribution, the angle multiplexing is straightforward. For example, by shifting each image to various positions within the angular spectrum through multiplication with plane waves of specific propagation angles corresponding to their desired positions, and then summing the resulting wavefronts, a complex amplitude distribution is generated to achieve the desired functionality. However, it becomes challenging for both polarization and angle multiplexing due to the constraint of the same amplitude for the two polarization channels (the same $D$). Moreover, this simple plane-wave summation method can lead to very low amplitude distribution and efficiency. To overcome these obstacles, we propose a gradient-based optimization method, as shown in Fig. 1b. Here, the three independent parameters $D$, $\alpha_1$, and $\alpha_2$ are used as inputs for optimization. The optical images observed at angles ($\theta_{xi}$, $\theta_{yj}$) with numerical aperture NA are:

$$H_{ij}^k = F^{-1}\left\{F\left[De^{i\alpha_k}(x,y)\right]T_{ij}(f_x,f_y)\right\} \qquad (3)$$

where $F$ and $F^{-1}$ represent the Fourier transformation and inverse Fourier transformation, respectively, $f_x$ and $f_y$ are the coordinates of the angular spectrum, and $T_{ij}(f_x, f_y)$ is the transfer function, which is 1 if $(f_x - f_{xi})^2 + (f_y - f_{yj})^2 \leq NA^2$ and 0 otherwise, where $f_{xi} = \sin(\theta_{xi})$ and $f_{yj} = \sin(\theta_{yj})$. To avoid image crosstalk between adjacent observed angles, the distance between adjacent observation angles in the angular spectrum $\Omega$ is set to be greater than 2×NA. Then, the mean squared error between all the calculated optical images $H_{ij}^k$ and the designed targets $D_{ij}^k$ is served as the loss function. Given that the entire process is gradient-calculable, the input parameters ($D$, $\alpha_1$, and $\alpha_2$) can be optimized to minimize the loss and ultimately realize the desired functionalities.

As a demonstration, we chose 5×5 ($M$=5) different observation angles, totaling 25 images under each polarization channel. The incident wavelength was set at 780 nm, with other parameters of $\Omega = 0.18$ and NA = 0.08. Figure 2c shows the plot of the loss versus epoch, converging after approximately 300 epochs. The obtained amplitude distribution $D$ (Fig. 2d) reveals that most areas reached unity amplitude, with an average amplitude of 0.89, demonstrating high efficiency of the metasurface. In comparison, the plane-wave summation method designed for only one polarization channel yielded a varied distribution (Fig. 2e), with only an average amplitude of 0.11. Figure 2f shows the simulated optical images under 25 observed angles and two polarization channels, which align well with our designed targets. The voids with null intensity in some images are due to the singularity of the phase distributions, which is challenging to eliminate due to the multiple-objective and high-efficiency constraints.

For the experimental demonstration, we first converted $D$, $\alpha_1$, and $\alpha_2$ to $\varphi_x$, $\varphi_y$ and $\theta$. The phase shift values $\varphi_x$ and $\varphi_y$ are associated with the transverse dimensions (length and width) of the nanoblock. We conducted full-wave finite-difference time-domain (FDTD) simulations to create a database of amplitude and phase shifts as functions of the nanoblock's transverse dimensions for both $x$- and $y$-polarizations. Using this dataset, any combinations of $\varphi_x$ and $\varphi_y$ values ranging from 0 to $2\pi$ can be realized by appropriately selecting the transverse dimensions of the nanoblocks (see details in Supplementary Information).

We fabricated the crystal silicon metasurface featuring a height of 600 nm and a period of 250 nm using electron beam lithography (EBL) and reactive ion etching (RIE) processes, with the details shown in Method section. The scanning electron microscope (SEM) image of the metasurface sample is displayed in Fig. 3a. For the experimental measurement (Fig. 3b), a continuous spectrum laser source with specific wavelengths isolated using an acousto-optic tunable filter (AOTF) was used to generate the incident laser beam. The circular polarization was generated by a linear polarizer (LP) and a quarter waveplate (QWP), and then normally incident on the metasurface. The light scattered by the metasurface was collected by a 40×/0.75 objective and isolated with another pair of QWP and LP. To obtain optical images at various observed angles, a spatial filtering process was implemented in the Fourier plane. The Fourier plane of the objective lens, typically located inside its barrel, was externalized using Lenses 1 and 2, making it accessible for filtering. A variable iris was placed at the Fourier plane to act as the filter. The final image was focused using Lens 3 and captured on a CMOS camera. The measured intensity profiles across the two polarizations and 25 different observed angles are shown in Fig. 3c, closely matching our simulated images and capturing significant details of even the null void distributions.

Next, we considered full-parameter multiplexing, incorporating the wavelength dimension. In this scenario, the coefficients *A* and *B* are wavelength-dependent and correlated across different wavelengths, rendering them unsuitable as starting points. The parameters optimized as the input should be invariant under all conditions, which we chose as the geometric parameters of the nanoblocks: $W_1$, $W_2$, and $\theta$.

Figure 4a illustrates the optimization algorithm employed for full-parameter multiplexing. To acquire gradient information on the optical coefficients relative to geometric parameters at various wavelengths, we performed numerical calculations, sweeping the parameter *A* across different $W_1$ and $W_2$ values under various wavelengths. These data were then used as training inputs for a deep neural network (DNN) (Fig. 4b) to establish the relationships. The DNN had two input parameters ($W_1$ and $W_2$) and 2×*N* output neurons, with each wavelength represented by two neurons corresponding to the real and imaginary parts of *A*. Due to geometric symmetry of the nanoblock structure, the same DNN could be used for *B* by swapping $W_1$ and $W_2$ values. Since DNNs inherently provide gradient information, we could derive the gradients of *A* and *B* with respect to $W_1$ and $W_2$ values. Additional details about this DNN can be found in the Methods section and Fig. S2.

Utilizing the obtained coefficients *A* and *B*, we calculated the complex-amplitude distributions $C^k(\lambda_n)$ under the different polarization and wavelength channels according to Eq. 2. Subsequently, the optical images $H_{ij}^k(\lambda_n)$ under different observed angles, wavelengths, and polarization channels were calculated with Eq. 3. The mean squared error between the calculated images $H_{ij}^k(\lambda_n)$ and the designed targets $D_{ij}^k(\lambda_n)$ served as the loss function. The following algorithm was the same as that of the single-wavelength case in Fig. 2a.

As a demonstration, we set three wavelengths ($N=3$): 560 nm, 670 nm, and 780 nm. The other parameters were the same as in the single-wavelength case. The loss versus iteration number showed stable decay and convergence after approximately 400 iterations (Fig. 4c). Figure 4d displays the measured optical images under various conditions, where all the main features are clearly visible, aligning well with the simulation results (Fig. S3). All images remain distinct from one another with minimal crosstalk, demonstrating the algorithm's effectiveness for full-parameter multiplexing. The number of multiplexing channels reached 150, which, to our knowledge, is the highest number of multiplexing channels reported to date.

The independent control over the wavelength dimension enables full-color optical images. We selected three wavelengths—475 nm, 532 nm, and 633 nm—corresponding to the primary RGB colors. The incident laser had multi-channel wavelength capability, allowing it to emit all three wavelengths simultaneously. Given the varying losses of crystal silicon at different wavelengths, particularly the significant loss at 475 nm, the intensities of the incident laser at the three wavelengths were balanced to match those of the three components of RGB images. Figure 5 shows our measured full-color optical images under the two polarization channels and 25 observed angles, where a continuous martial arts movement in the LCP polarization channel and a rotating windmill in the RCP polarization channel can be observed. By continuously moving the iris in the Fourier plane, we extracted the different optical images observed from different angles, creating a dynamic movie.

Since LCP and RCP can be controlled independently, this technique can serve as two images for binocular observation, potentially enabling the design and realization of stereoscopic full-color dynamic images in the future.

## Conclusions

In summary, we have successfully demonstrated a metasurface capable of multiplexing all three optical parameters of plane waves—polarization, wavelength, and angle—resulting in 150 distinct channels. This achievement sets a new standard for the number of multiplexing channels. The metasurface design employs a gradient-based optimization algorithm, allowing independent control with minimal cross-talk among channels. Experimental validation showed high efficiency and strong agreement with predictions, confirming the robustness of our design. Future research could explore the application of this technique to broader wavelength ranges and more varied polarization states, for example, utilizing the three polarization channels of the single layer metasurface[5]. The capability to independently control optical parameters paves the way for future applications in complex and dynamic optical systems and for the development of next-generation optical devices with unprecedented capabilities in imaging, sensing, and communication technologies.

## Methods

**Metasurface fabrication.** To begin, a commercial silicon-on-insulator (SOI) wafer with an initial device layer thickness of 1200 nm was bonded to a glass substrate through adhesive wafer bonding and deep reactive ion etching (DRIE). The device layer thickness was then reduced to 600 nm using inductively coupled plasma (ICP) etching. Next, a 300 nm layer of hydrogen silsesquioxane (HSQ) was spin-coated onto the substrate at 4000 revolutions per minute and baked on a hot plate at 90 °C for 5 minutes. A 30 nm aluminum layer was thermally evaporated onto the HSQ to serve as a charge dissipation layer. Electron beam lithography at 30 kV was used to define the metasurface pattern on the HSQ layer. Following the exposure, the aluminum layer was etched away with a 5% phosphoric acid solution, and the resist was developed with tetramethylammonium hydroxide (TMAH) for 2 minutes at room temperature (25 °C). The pattern was then etched into the silicon using inductively coupled plasma-reactive ion etching (ICP-RIE). Finally, the samples were immersed in a 10% hydrofluoric (HF) acid solution for 15 seconds to remove the residual HSQ mask, followed by cleaning with deionized water and drying with nitrogen gas.

**DNN.** To establish the relationship between optical coefficients and geometric parameters, we utilized a fully connected linear DNN with six layers, each containing 200 neurons. The input layer had two neurons ($W_1$ and $W_2$), while the output layer comprised 2×N neurons, representing the real and imaginary parts of the optical coefficient $A$ at N wavelengths. A rectified linear unit (ReLU) activation function was applied between each layer. For training data, FDTD simulations were conducted to obtain the optical coefficient A across 36 linearly spaced values (from 60 nm to 200 nm) for $W_1$ and $W_2$, generating 1296 datasets. Due to symmetry, optical coefficients $B$ were derived by swapping the mappings of $W_1$ and $W_2$. We then interpolated these 1296 data points to expand the dataset to 20,000 points. The DNN was trained with a batch size of 5000 using the Adam optimizer with a learning rate of 0.01. The dataset was divided into training and testing sets, with 80% for training and 20% for testing. Fig. S2 illustrates the train and test loss, along with the comparison between predicted and true values for the 3-wavelength (560 nm, 670 nm and 780 nm) scenarios.

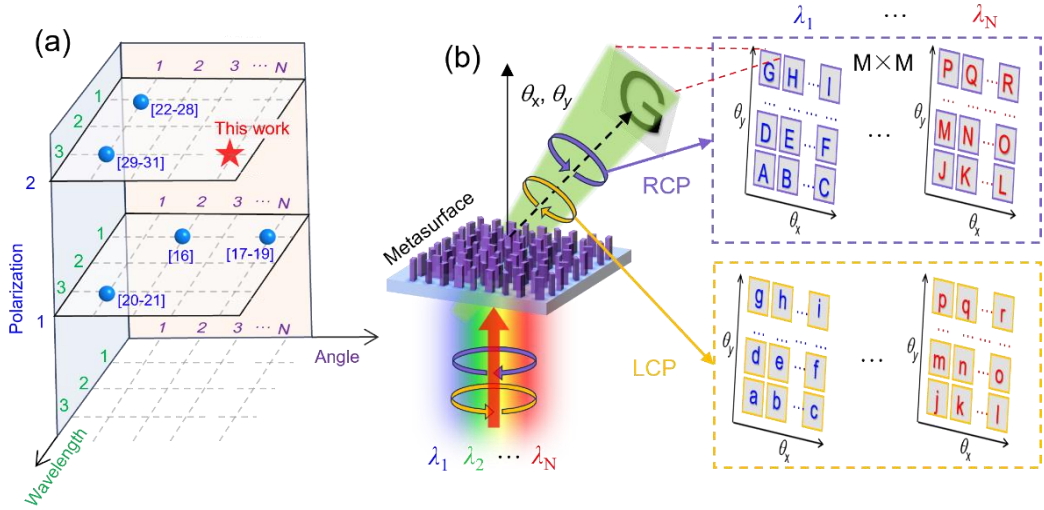

**Fig. 1. Full-parameter optical multiplexing with metasurface.** (a) A summary of optical multiplexing of wavelength, polarization, and angle in the literature. The blue spheres indicate the realized multiplexing with the reference numbers illustrated beside them. Our work is indicated by a red pentacle, achieving the full-parameter multiplexing. (b) A schematic view of full-parameter multiplexing demonstration. The metasurface enables the multiplexing of two polarization (LCP and RCP), $N$ wavelength channels ($\lambda_1$, $\lambda_2$, ..., $\lambda_N$), and $M \times M$ observed angles, resulting in a total of $2M^2N$ multiplexing channels.

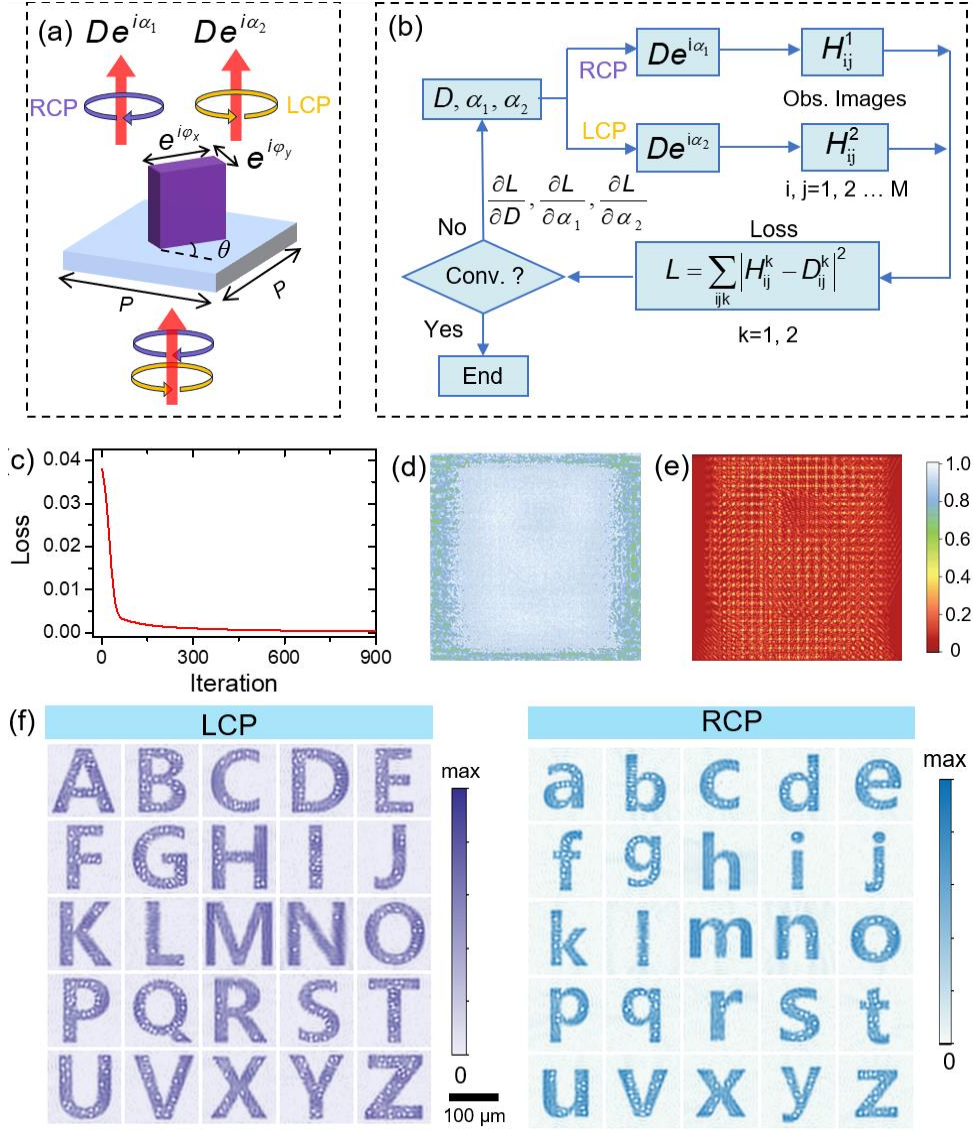

**Fig. 2. Polarization and angle multiplexing with single wavelength.** (a) Schematic of a nanoblock nanostructure for circular polarization manipulation. At single wavelength, the nanoblock imposes pure phase modulation of $e^{i\varphi_x}$ and $e^{i\varphi_y}$ along the fast and slow axes, respectively, resulting in the same amplitude $D$ and independent phases $\alpha_1$ and $\alpha_2$ for the LCP and RCP polarization channels. The period of the nanoblock is $P=250$ nm and its rotation angle is $\theta$. (b) The gradient-base optimization algorithm for polarization and angle multiplexing. The optimized parameters are the three parameters of $D$, $\alpha_1$, and $\alpha_2$. (c) Loss value as a function of the iteration number. (d) Distribution of the amplitude $D$. (e) Distribution of the amplitude of the complex-amplitude with plane-wave summation method designed

for only one polarization channel. (f) Simulated optical images of 25 observed angles under LCP (left) and RCP (right) polarizations.

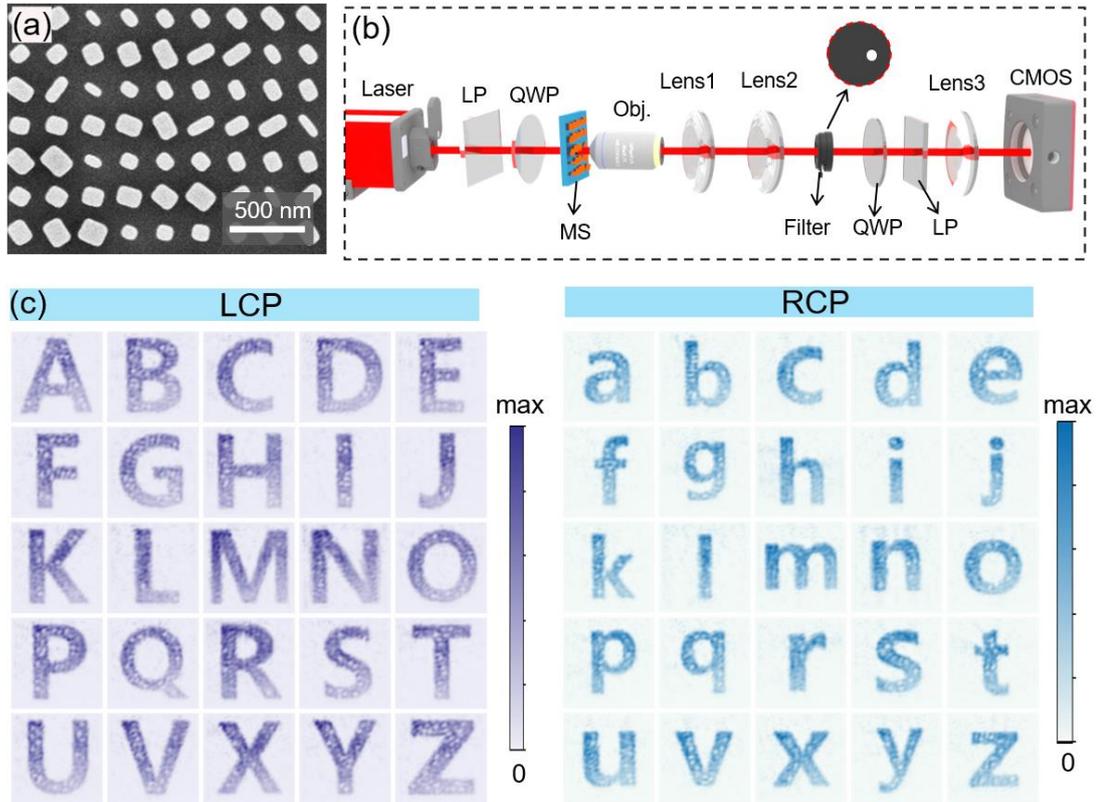

**Fig. 3. Experimental demonstration of polarization and angle multiplexing.** (a) Scanning electron microscopy (SEM) images of the fabricated metasurface (partial view). (b) Optical setup for measurement. The focal lengths of Lenses 1, 2 and 3 are 10 cm, 20 cm and 20 cm, respectively. (LP: linear polarizer, QWP: quarter waveplate). (c) Measured optical images of 25 observed angles under LCP (left) and RCP (right) polarizations.

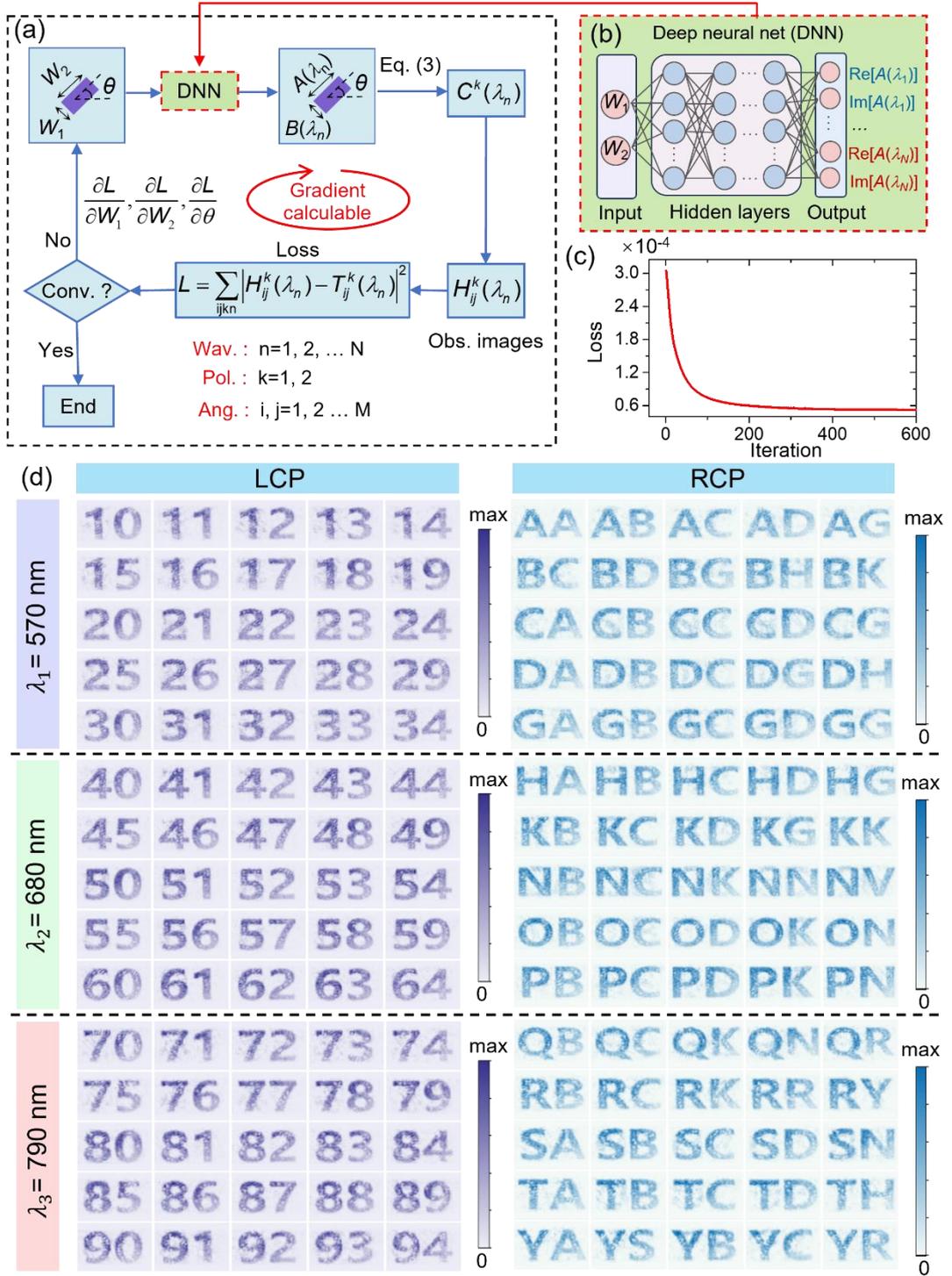

**Fig. 4. Design of full-parameter multiplexing of wavelength, polarization and angle.** (a) Schematic of the gradient-base optimization algorithm. The optimized parameters are the three geometric parameters of nanoblocks: $W_1$, $W_2$ and $\theta$. A DNN is used to derive the gradients of $A$ and $B$ concerning $W_1$ and $W_2$ values. (b) Schematic of the DNN for establishing the relationship between optical

coefficient *A* and geometric parameters. The DNN is a fully connected linear network with the input layer having two neurons ($W_1$ and $W_2$), and the output layer has 2×*N* neurons, corresponding to the real and imaginary parts of optical coefficient *A* at *N* wavelengths. This DNN can be used for *B* by swapping $W_1$ and $W_2$ values due to symmetry. (c) Loss value as a function of the iteration number. (d) Measured optical images multiplexed within 25 observed angles, two polarization (LCP and RCP) and three wavelengths (560 nm, 670 nm, and 780 nm).

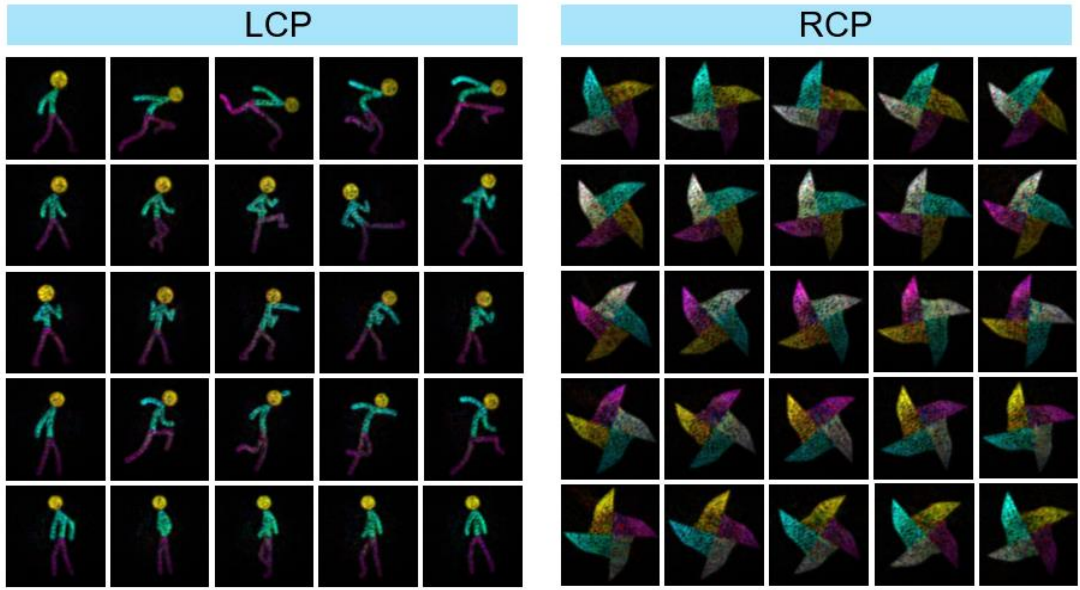

**Fig. 5. Measured full-color RGB images multiplexed in polarization and angle channels.** The images show a continuous martial arts movement in the LCP polarization channel and a rotating windmill in the RCP polarization channel.